\documentclass{aa}

\usepackage{graphicx}
\usepackage{txfonts}
\usepackage{hyperref}

\begin{document}

   \title{The first near-infrared high-resolution {\'e}chelle spectroscopy\\of the outflow in NGC 4151}

   \subtitle{A study of the clouds covering the Eye of Sauron}

   \author{F. G. Saturni\inst{1}
          \and
          R. Middei\inst{1,2}
          \and
          H. Landt\inst{3}
          \and
          V. D'Elia\inst{4}
          \and
          F. La Franca\inst{5}
          \and
          M. Perri\inst{1,4}
          \and
          E. Piconcelli\inst{1}
          }

   \institute{INAF -- Osservatorio Astronomico di Roma, Via Frascati 33, I-00078 Monte Porzio Catone (RM), Italy\\
              \email{francesco.saturni@inaf.it}
         \and
            Center for Astrophysics | Harvard \& Smithsonian, 60 Garden Street, Cambridge MA 02138, USA
         \and
             Centre for Extragalactic Astronomy, Department of Physics, Durham University, South Road, Durham DH1 3LE, UK
         \and
             ASI -- Space Science Data Center, Via del Politecnico snc, I-00133 Roma, Italy
         \and
             Dip. di Matematica e Fisica, Universit{\`a} degli Studi ``Roma Tre'', Via della Vasca Navale 84, I-00146 Roma, Italy
             }

   \date{Received 30 Oct 2025; accepted 24 May 2026}
 
  \abstract
  {We present the first high-resolution near-infrared spectroscopy of the nucleus of the nearby, well-known Seyfert galaxy NGC 4151 (the ``Eye of Sauron''). Past studies of this sources have revealed that it exhibits a variable absorption feature associated with the He {\scriptsize I} $\lambda$10,830 \AA\ emission line, potentially indicative of obscuration events affecting the central engine. Here, we take advantage of the {\it IRTF}/iSHELL and {\it TNG}/GIANO-B spectrographs to observe this feature with unprecedented spectral resolution ($\lambda/\Delta\lambda > 50,000$), being able to study in detail the structure of the absorption trough and its variations over a time span of $\sim$700 days. In order to infer a connection between the He {\footnotesize I} absorption variability and that of the X-ray ionising continuum, we also analyse the publicly available data collected by the {\it Swift}-XRT instrument over the same period of time, unveiling a potential driving mechanism in the changes of the outflow ionisation state due to the X-ray flux variations. We also derive outflow physical parameters -- $\dot{M}_{\rm out} \lesssim 10^{-2}$ M$_\odot$ yr$^{-1}$, $r_{\rm out} \sim 3$ pc, $v_{\rm max} \sim 1000$ km s$^{-1}$ -- that are in line with those of comparable ionised winds found in similar targets, where the outflow is not powerful enough to trigger a significant AGN feedback ($\dot{E}_{\rm kin}/L_{\rm bol} \sim 0.001$\%). Such findings point at a scenario in which a dusty and clumpy outflow that obscures NGC 4151 up to galactic scales responds to changes in the ionising X-ray flux, similarly to what happens in quasars with broad absorption lines and Seyferts with multi-phase outflows such as NGC 5548.}


   \maketitle

\section{Introduction}

Active galactic nuclei (AGN) are energetic processes located at the centres of galaxies \citep{lyn69}, and powered by material accreting onto a supermassive black hole \citep[SMBH; e.g.,][]{sha73}. AGN outflows limit the rate of SMBH growth and can significantly influence star formation in their host galaxy by depositing their kinetic energy in the surrounding environment \citep{spr05,mar18}. The outflowing material, if directly located along our line-of-sight to the AGN, appears as blue-shifted absorption lines, offset from the rest-frame wavelength by velocities sometimes exceeding 10\% the speed of light. Such absorption features have been observed at several wavelengths, especially as broad absorption lines \citep[BALs; e.g.,][]{lyn67,wey91} in near-IR (NIR), optical and UV, and as ultra-fast outflows \citep[UFOs; e.g.,][]{tom10,vie22} in the X-rays.

The outflowing gas is often found in various ionisation states, implying a corresponding variety of physical phases and ionising environments \citep[e.g.,][]{cic18}; in particular, high-column density outflows are predicted in AGN that provide the highest kinetic energies, thus making such sources invaluable test beds for galaxy evolution models \citep[see e.g.][and refs. therein]{ham19}. Atomic helium He {\footnotesize I} in the $2^3S$ state is metastable, and, due to its relatively low abundance that makes it less prone to saturation, can thus be used to study high-column density outflows. In AGN outflows, metastable He {\footnotesize I} is almost entirely formed from He {\footnotesize II} recombination. The transitions of metastable He {\footnotesize I} form an emission triplet at wavelengths 3189 \AA\ (UV), 3890 \AA\ (optical) and 1.08 $\mu$m (NIR), the NIR component being the strongest. The effectiveness of using metastable He {\footnotesize I} as a mean to understand AGN outflows was first demonstrated by \citet{lei11}: by analysing the He {\footnotesize I} properties in the FBQS J1151+3822 quasar, they put important constraints on the location of the absorber based on the ionisation parameter, on the number and column densities $N_{\rm HeI}$, and on the spatial extension of the outflow -- i.e. the covering fraction. However, to date, no high-resolution near-IR spectroscopy exists on any AGN known to harbour He {\footnotesize I} absorbers.

At variance with UV and X-ray wavelengths, where studies of outflows at high spectral resolution are common \citep[e.g.,][]{ell99,kas06,ara08,ara20,che21}, only a few AGN are known to exhibit He {\footnotesize I} $\lambda$10,830 absorbers \citep{lei11,lei14,ji15,wil16,liu16,zha17,lan19,wil21,liu21,mao22} due to the need for them to form in environments that permit the production of He {\footnotesize II} ions \citep{wil21}, and almost all of them are weak \citep[$N_{\rm HeI} < 10^{15}$ cm$^{-2}$; e.g.,][]{lei11}. In this framework, instruments like iSHELL \citep[Immersion Grating {\'E}chelle Spectrograph;][]{ray22} on the NASA {\it Infrared Telescope Facility} ({\it IRTF}, Hawai'i Islands, the USA) and GIANO-B on the {\it Telescopio Nazionale Galileo} ({\it TNG}, Canary Islands, Spain) provide a unique opportunity to study the NIR He {\footnotesize I} absorbers.

NGC 4151 \citep{may34} at $z = 0.0033$ -- commonly known as the ``Eye of Sauron'' for its visual appearance -- is the brightest and most nearby AGN exhibiting such a feature in its NIR spectrum, which has also been found to be variable \citep{wil16}. To date, the only other science case of an AGN showing a variable He {\footnotesize I} absorber is NGC 5548, as recently discovered by \citet[see their fig. 1]{lan19}. Moreover, \citet{wil21} reported on the discovery of both He {\footnotesize I} broad and narrow absorption in NGC 5548, associating the former with the obscurer itself and the latter with known UV components of the so-called `warm absorber' (WA). For such reasons, we aim at studying this feature with unprecedentedly high spectral resolution using iSHELL and GIANO-B data, in order to determine for the first time the nature of the obscuring/absorbing matter also in the nucleus of NGC 4151.

The paper is organised as follows: in Sect. \ref{sec:obsred}, we present the observations and the adopted data reduction procedures; in Sect. \ref{sec:specdec}, we describe the fitting procedures used to decompose the NGC 4151 NIR spectra; in Sect. \ref{sec:xraydat}, we outline the X-ray data reduction procedure and correlate the NGC 4151 X-ray spectral parameters with those of its He {\footnotesize I} absorption; finally, we discuss our results in Sect. \ref{sec:disc} and summarise our work in Sect. \ref{sec:conc}. Throughout the text, we adopt a $\Lambda$-CDM cosmology with $H_0 = 70$ km s$^{-1}$ Mpc$^{-1}$, $\Omega_{\rm M} = 0.3$ and $\Omega_\Lambda = 0.7$.

\section{Observations and data reduction}\label{sec:obsred}

The {\it IRTF} observations were performed in contiguous dates, on 2020 May 18/19 and 2021 Jan 31/Feb 01 respectively, in the framework of a monitoring campaign aimed at characterising the time evolution of the NGC 4151 absorption properties (PI: H. Landt). The acquisitions were composed of pairs containing both a high-resolution iSHELL spectrum in {\'e}chelle mode with the $J0$ setting to cover the He {\footnotesize I} spectral interval $\lambda\lambda$1.062--1.165 $\mu$m in the observer frame with a resolving power $\lambda/\Delta\lambda \sim 75,000$, and a medium-resolution one ($\lambda/\Delta\lambda \sim 2000$) taken with the SpeX spectrograph \citep{ray03} over the range $\lambda\lambda$0.7--2.5 $\mu$m for a precise broadband modeling of the NGC 4151 NIR continuum emission. The GIANO-B {\'e}chelle spectrum was taken on 2022 May 20 (PI: R. Middei) with the goal of detecting the He {\footnotesize I} $\lambda$10,830 feature in the $\lambda\lambda$1.065--1.105 $\mu$m spectral interval ($38^{\rm th}$ and $39^{\rm th}$ {\'e}chelle orders) at $\lambda/\Delta\lambda \sim 50,000$; for the same purpose of broadband continuum modeling, also a low-resolution NICS \citep[Near Infrared Camera Spectrometer;][]{baf01} spectrum was acquired on 2022 Jun 15 with the $IJ$ grism ($\lambda\lambda$0.85--1.45 $\mu$m spectral range, $\lambda/\Delta\lambda \sim 500$). In Tab. \ref{tab:jobs}, we report the relevant informations of each observation.

\begin{table}
\centering
 \caption{Journal of observations. The signal-to-noise ratios $S/N$ of each resulting spectrum are computed according to the prescriptions listed in \citet{ros12}.}
 \label{tab:jobs}
 \resizebox{\linewidth}{!}{
 \LARGE
 \begin{tabular}{ccccccc}
 \hline
 \hline
 \multicolumn{7}{l}{ }\\
 Date & Instrument & $\lambda$ range & Exp. time & Airmass & Slit size & $S/N$\\
 (YYYY MMM DD) & & ($\mu$m) & (s) & & (arcsec$^2$)\\
 \multicolumn{7}{l}{ }\\
 \hline
 \multicolumn{7}{l}{ }\\
     2020 May 18 & {\it IRTF}/iSHELL & $1.06 - 1.16$ & 4197 & 1.27 & $5 \times 0.75$ & 7.1\\
     2020 May 19 & {\it IRTF}/SpeX & $0.69 - 2.57$ & 359 & 1.06 & $15 \times 0.3$ & 35.3\\
     2021 Jan 31 & {\it IRTF}/iSHELL & $1.06 - 1.16$ & 2099 & 1.20 & $5 \times 0.75$ & 6.4\\
     2021 Feb 01 & {\it IRTF}/SpeX & $0.69 - 2.57$ & 956 & 1.10 & $15 \times 0.3$ & 53.5\\
     2022 May 20 & {\it TNG}/GIANO-B & $1.00 - 1.12$ & 4200 & 1.14 & ({\'e}chelle) & 26.9\\
     2022 Jun 15 & {\it TNG}/NICS & $0.86 - 1.45$ & 1600 & 1.10 & $4 \times 1$ & 16.8\\
 \multicolumn{7}{l}{ }\\
 \hline
 \end{tabular}
 }
 \end{table}

\begin{figure*}[htbp]
    \centering
    \includegraphics[scale=.5]{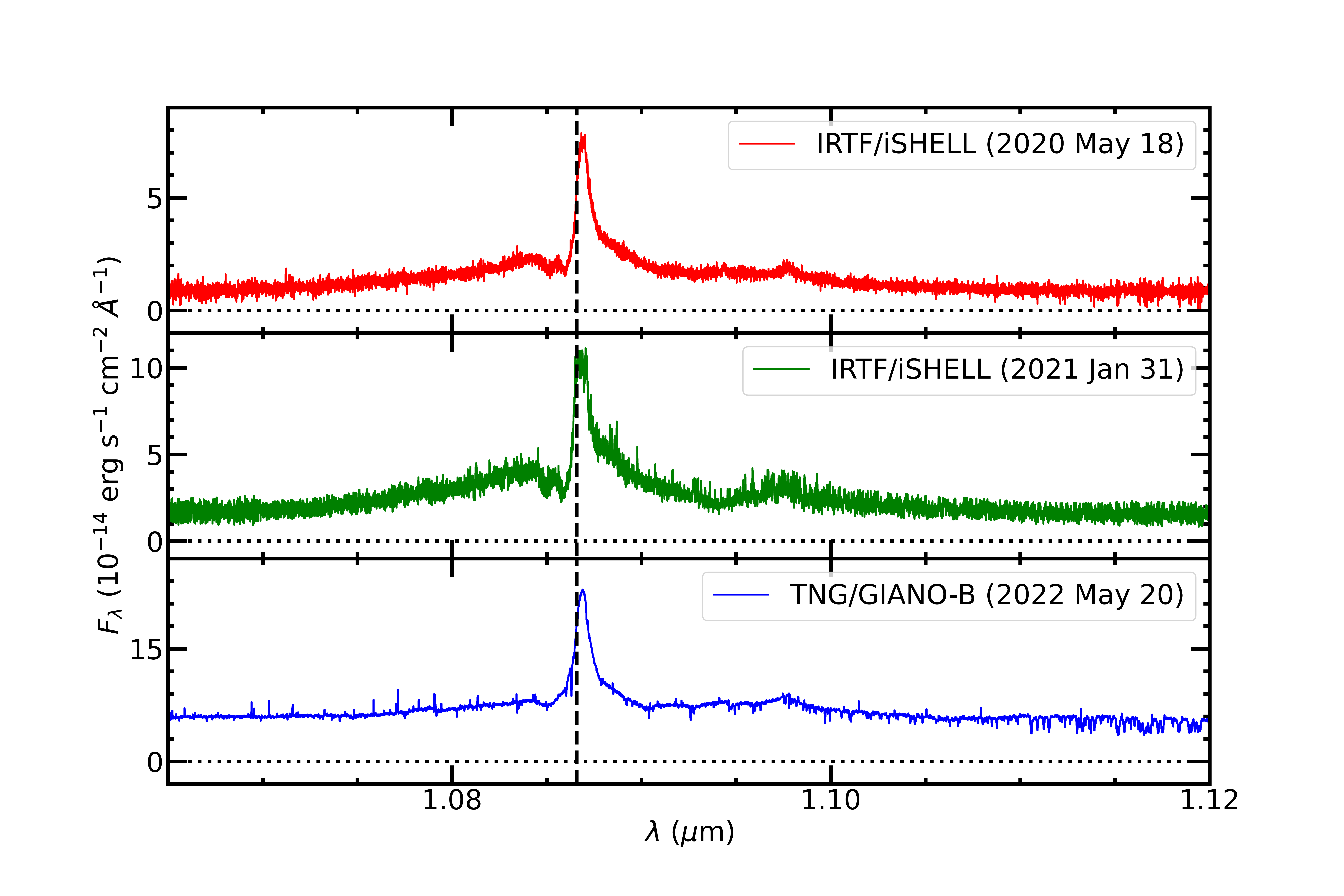}
    \caption{NGC 4151 high-resolution spectra around the He {\scriptsize I} $\lambda$10,830 \AA\ emission feature in the observer frame. In each panel, the instrument and epoch of acquisition is indicated ({\it see legends}), along with both the zero-flux level ({\it dotted lines}) and the position of the He {\scriptsize I} transition in the observer frame ({\it dashed line}). The different flux levels are due to the lack of an inter-calibration among the spectra.}
    \label{fig:hires}
\end{figure*}

Each {\it IRTF} spectrum has been reduced using the appropriate pipelines, namely {\ttfamily Spextool v4.1} \citep{cus04} for SpeX and {\ttfamily v5} for iSHELL. Similarly, the NICS observation has been reduced through standard {\footnotesize IRAF} \citep{tod86} and {\footnotesize MIDAS} routines \citep{war91}. The GIANO-B exposure was initially reduced with the online data reduction software \citep[DRS;][]{har18}; however, this resulted in an incorrect reconstruction of the He {\footnotesize I} emission profile, in which the broad component was disappearing due to the interpolation and extraction procedure of the {\'e}chelle orders. To overcome this issue, we first visually compare the wavelength ranges of the GIANO-B {\'e}chelle orders with the iSHELL spectra: in this way, we check that the He {\footnotesize I} feature is not extending beyond the $38^{\rm th}$ and $39^{\rm th}$ orders. Therefore, we decide to perform the last steps of the GIANO-B data reduction with a custom {\ttfamily Python} routine that allows us to correctly extract the He {\footnotesize I} spectral region through the interpolation of the immediately adjacent orders, in which the residual signal of the He {\footnotesize I} is negligible. We present the final high-resolution spectra in Fig. \ref{fig:hires}: a visual inspection already reveals a certain degree of variability in both shape and flux level of the He {\footnotesize I} absorption.

\section{Spectral decomposition and analysis}\label{sec:specdec}

The effect of differential light absorption by an intervening medium with optical depth $\tau_\lambda$ on a spectrum composed by a continuum $C_\lambda$ and a system of emission-line components $L_\lambda$ can be mathematically written as \citep[e.g.,][]{ham99,tre13}:
\begin{equation}\label{eqn:absfeat}
    F_\lambda = \left(
    1 - f_C
    \right) C_\lambda + \left(
    1 - f_L
    \right) L_\lambda + \left(
    f_C C_\lambda + f_L L_\lambda
    \right) e^{-\tau_\lambda}
\end{equation}
where $f_C$ and $f_L$ are the outflow covering factors of the continuum and line emission regions, respectively. Since our spectral section of interest is dominated by the emission from regions that are likely more spatially extended than the wind -- namely, the broad-line region (BLR), the narrow-line region (NLR) and a dusty torus surrounding the central engine \citep{wil16} -- we expect that both $f_C$ and $f_L$ are $\lesssim$1 \citep{tre13}. However, disentangling the relative fractions of coverage on these various emitting regions is nontrivial for the case of NGC 4151, since the He {\footnotesize I} trough insists on all the emission components (see Fig. \ref{fig:hires}). Therefore, in the following we assume $f_C = f_L = 1$ for simplicity; such an approach is commonly adopted in the literature \citep[e.g.,][]{cap11}, its major effect being an increment of order unity of the measured absorption intensities\footnote{The assumption of total coverage for the outflow may lead to more severe underestimates of the absorption intensity in case of very different -- and varying -- partial coverages for the various AGN emitting components \citep[e.g.,][]{ham98,gan99,ara05,gre23}; in such cases, whose analysis is beyond the scope of this paper, all the He {\scriptsize I} absorption strengths computed in the following should be interpreted as lower limits on the real values.} \citep[see][and refs. therein]{tre13}.

\begin{figure*}[htbp]
    \centering
	\includegraphics[scale=.7]{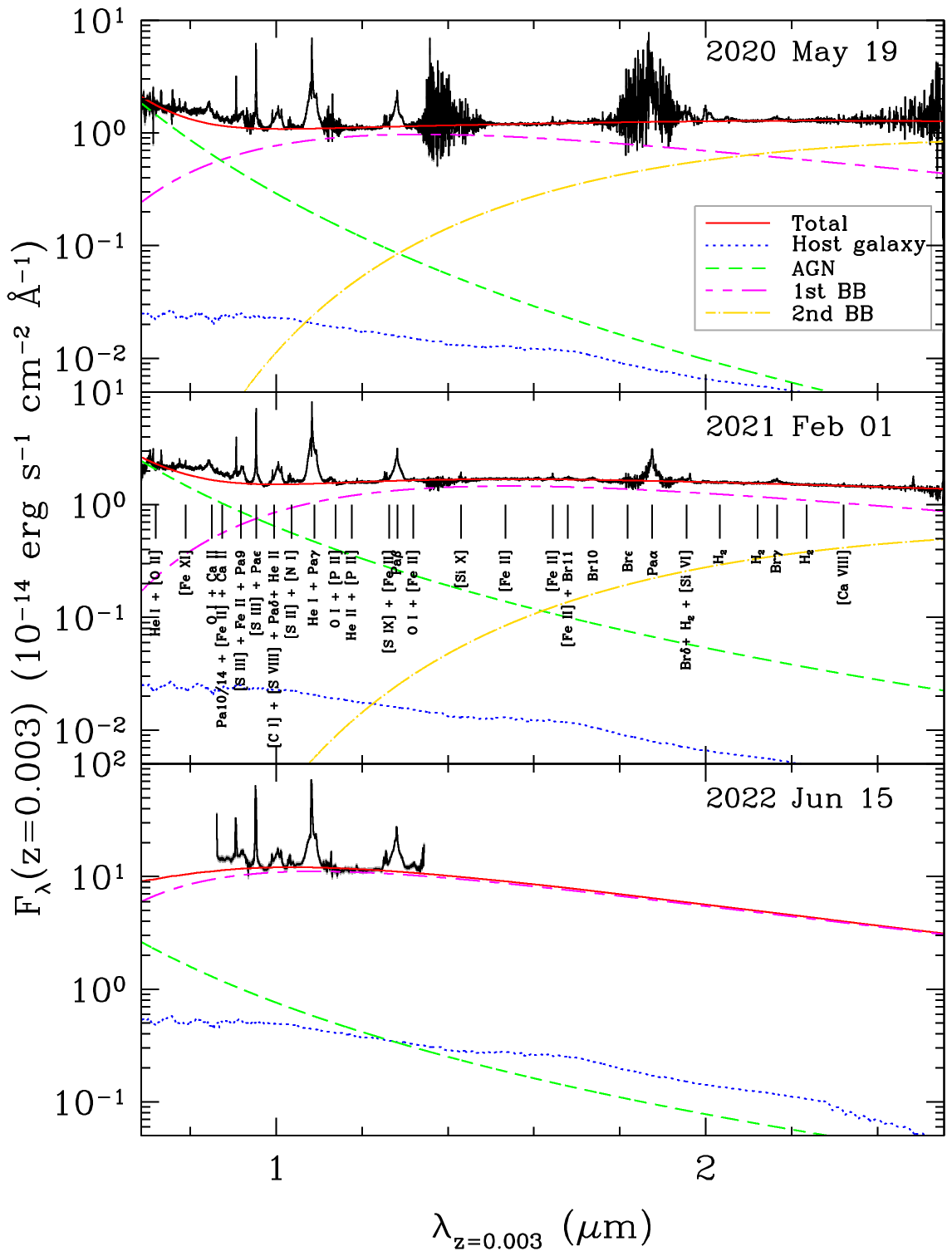}
    \caption{{\itshape Top panel:} rest frame NGC 4151 low-resolution spectrum taken with {\itshape IRTF}/SpeX on 2020 May 19th. {\itshape Middle panel:} spectrum taken with {\itshape IRTF}/SpeX on 2021 February 1st. {\itshape Bottom panel:} spectrum taken with {\itshape TNG}/NICS on 2022 June 15th. In all panels, the best fit to the continuum emission ({\itshape red solid line}) is shown superimposed to the data ({\itshape black solid line}) along with the host-galaxy template spectrum ({\itshape blue dotted line}), the AGN power law ({\itshape green short-dashed line}) and the two BB components ({\itshape magenta long-short-dashed line and yellow dot-dashed line}). The major IR emission lines \citep{lan08} are labelled. As per the high-resolution spectra, the different flux levels are due to the lack of an inter-calibration procedure.}
    \label{fig:ngc4151irspec}
\end{figure*}

\subsection{The continuum fit}

We first model the NGC 4151 continuum emission with an empirical combination of emission components, exploiting the line-free intervals of the low-resolution spectra. According to \citet{wil16}, the continuum spectrum can be fitted by a combination of host-galaxy stellar emission, power-law (PL) emission from the AGN accretion disk, and black-body (BB) emission by the torus. The host contribution has been derived by combining the Sa- and Sb-type emission templates from the SWIRE library of galaxy spectral energy distributions\footnote{Available at {\ttfamily http://www.iasf-milano.inaf.it/$\sim$polletta/\\templates/swire\_templates.html}.} \citep[SEDs;][]{pol07}. We note that the 2020 and 2021 epochs were acquired with a narrower slit ($0''.3$) with respect to the epochs analysed in \citet[$0''.8$]{wil16}; this makes the host-galaxy contribution negligible, thus causing the fitting procedure to zero out its flux normalisation. To overcome this issue, for these two epochs we fix the host-galaxy contribution to the average of the estimated fluxes derived from the {\itshape HST} observations analysed in \citet{lan11}; such fluxes have been adequately rescaled to the different slit areas ($0''.3 \times 15''$ and $1'' \times 4''$, respectively) used to acquire the SpeX and NICS spectra, with respect to the {\itshape HST} apertures ($3'' \times 4''.8$ to $3'' \times 8''.4$).

In performing the continuum fit of the 2020 and 2021 epochs with the emission model by \citet{wil16}, we find that, at variance with their results, the continuum level in the wavelength range 0.7--0.9 $\mu$m cannot be well reproduced, unless a second BB component peaking at a shorter wavelength is added to the fit in both epochs to correctly model the observed flux level blueward of the Pa$\epsilon$ and Pa$\delta$ transitions. According to an $F$-test, the inclusion of this additional component is statistically significant with respect to the single BB model for both epochs at 6.4$\sigma$ and 7.6$\sigma$ confidence level, respectively.

Due to the more limited wavelength extension of the NICS spectrum (0.85--1.35 $\mu$m) with respect to the SpeX ones (0.7--2.55 $\mu$m), the fit with two BB components cannot be constrained in the 2022 epoch. Therefore, we report the temperatures of the BB emitters derived for the double-BB best-fit model only in Tab. \ref{tab:ngc4151bfp} for the 2020 and 2021 epochs only, along with the PL slopes of the AGN continuum. On average, the adopted empirical BB emissions have $T_{\rm BB}^{(1)} = 1900 \pm 100$ K and $T_{\rm BB}^{(2)} = 900 \pm 50$ K respectively; such values are broadly consistent with the highest temperatures found by \citet{lyu21} in the modeling of the NGC 4151 SED with AGN emission components from warm dust (see their figure 18). The best-fit continuum models of the NGC 4151 low-resolution spectra at each epoch are shown in Fig. \ref{fig:ngc4151irspec}; in the following, we use this model only to subtract the underlying continuum emission from the high-resolution {\it IRTF}/iShell and {\it TNG}/GIANO-B He {\footnotesize I} spectra.

\subsection{The emission-line fit}

\begin{figure}[htbp]
    \centering
	\includegraphics[scale=.5]{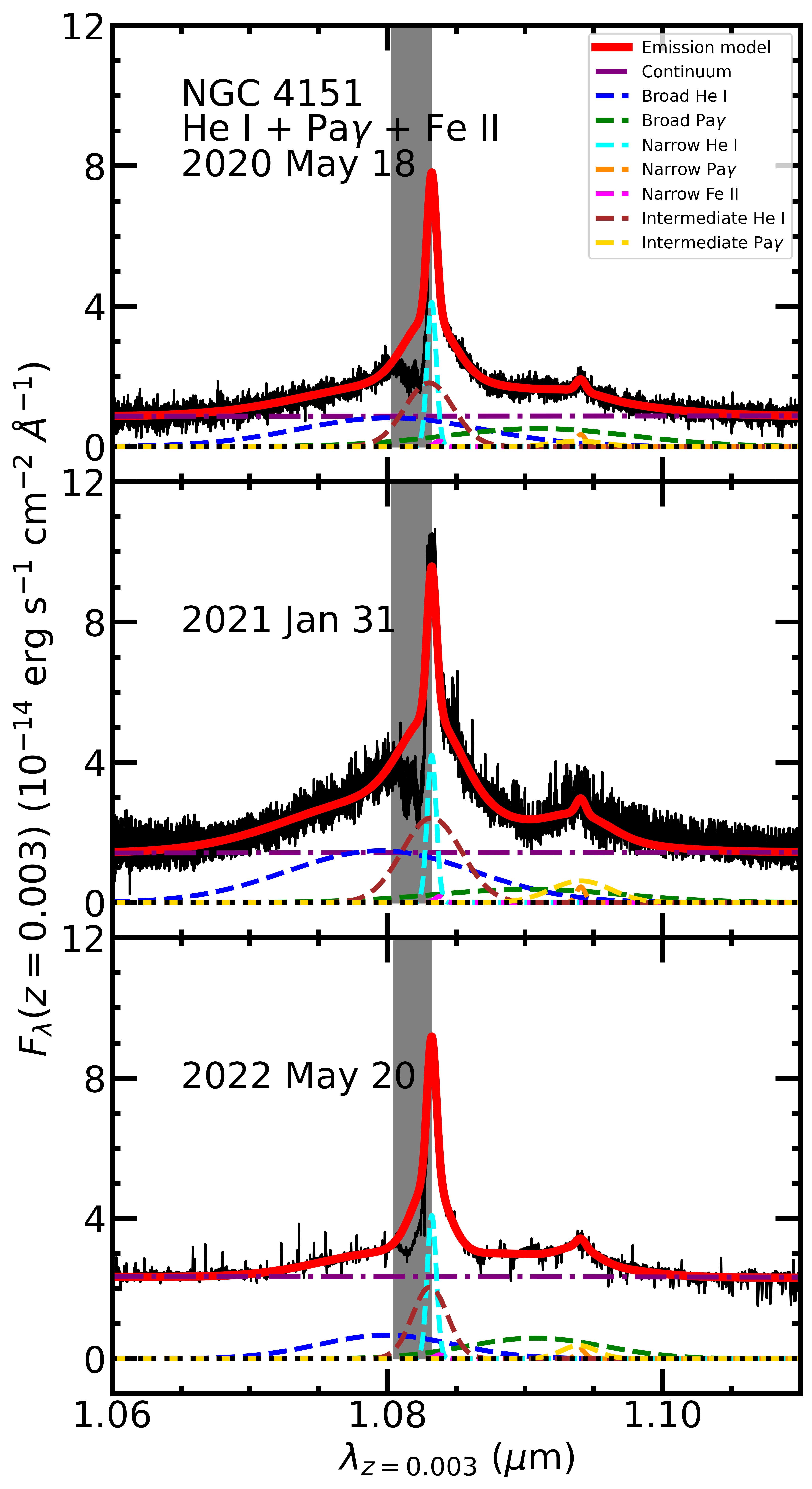}
    \caption{Rest-frame best-fit profiles of the He {\scriptsize I}$+$Pa$\gamma$ blended emission around 1.08 $\mu$m. {\itshape Top panel:} 2020 May 18th epoch. {\itshape Middle panel:} 2021 January 31st epoch. {\itshape Bottom panel:} 2022 May 20th epoch. In all panels, the total unabsorbed emission profile ({\itshape red solid line} is shown superimposed to the data ({\itshape black solid line}), along with the continuum flux level ({\itshape purple dot-dashed line}) and the emission components adopted for the line decomposition ({\itshape dashed lines}). In addition, both the zero-level flux ({\it black dotted line}) and the masked region ({\it grey band}) are indicated.}
    \label{fig:ngc4151he1}
\end{figure}

In all epochs, the He {\footnotesize I} $\lambda$10,830 \AA\ emission line shows an absorption complex located blue-ward of the systemic peak wavelength that can be interpreted as due to an outflow with moderate velocity. In order to evaluate the physical parameters of this absorber, we perform an empirical spectral decomposition of the He {\footnotesize I}$+$Pa$\gamma$ blend based on multiple Gaussian emission components, with at least two of them -- a broad (FWHM $\gtrsim$ 2000 km s$^{-1}$) and a narrow one (FWHM $\lesssim$ 500 km s$^{-1}$) -- for every allowed transition. Such a model is similar to that performed by \citet{wil16} on a 2015 spectrum of NGC 4151, which is free of major absorption features; at variance with them, we also include a narrow Fe {\footnotesize II} contribution located at $\lambda \sim 1.085$ $\mu$m \citep{lan08}. First, we fit the [S {\footnotesize III}] $\lambda$9531 emission line on the low-resolution SpeX and NICS spectra to infer the full width at half maximum (FWHM) of the narrow emission components. To this aim, we subtract the continuum emission shown in Fig. \ref{fig:ngc4151irspec} from the corresponding data; then, we model the line profile with a single Gaussian that is left completely free to vary in each epoch. Then, we compute the intrinsic FWHM of the narrow-line emissions by removing the systematic width increase due to the spectral resolution $\Delta v_{\rm sys}$:
\begin{equation}\label{eqn:fwhmcorr}
    \resizebox{.9\hsize}{!}{${\rm FWHM}_{\rm int} = \sqrt{{\rm FWHM}_{\rm obs}^2 - \Delta v_{\rm sys}^2} = \sqrt{{\rm FWHM}_{\rm obs}^2 - \left(
    \frac{c}{\lambda/\Delta\lambda}
    \right)^2}$}
\end{equation}
For all spectra, we get compatible results within 1$\sigma$ uncertainties. Therefore, we adopt FWHM$_{\rm int} = 220 \pm 20$ km s$^{-1}$ as the common value for the width of narrow emission components in the He {\footnotesize I}$+$Pa$\gamma$ complex.

\begin{figure}[htbp]
    \centering
	\includegraphics[scale=.5]{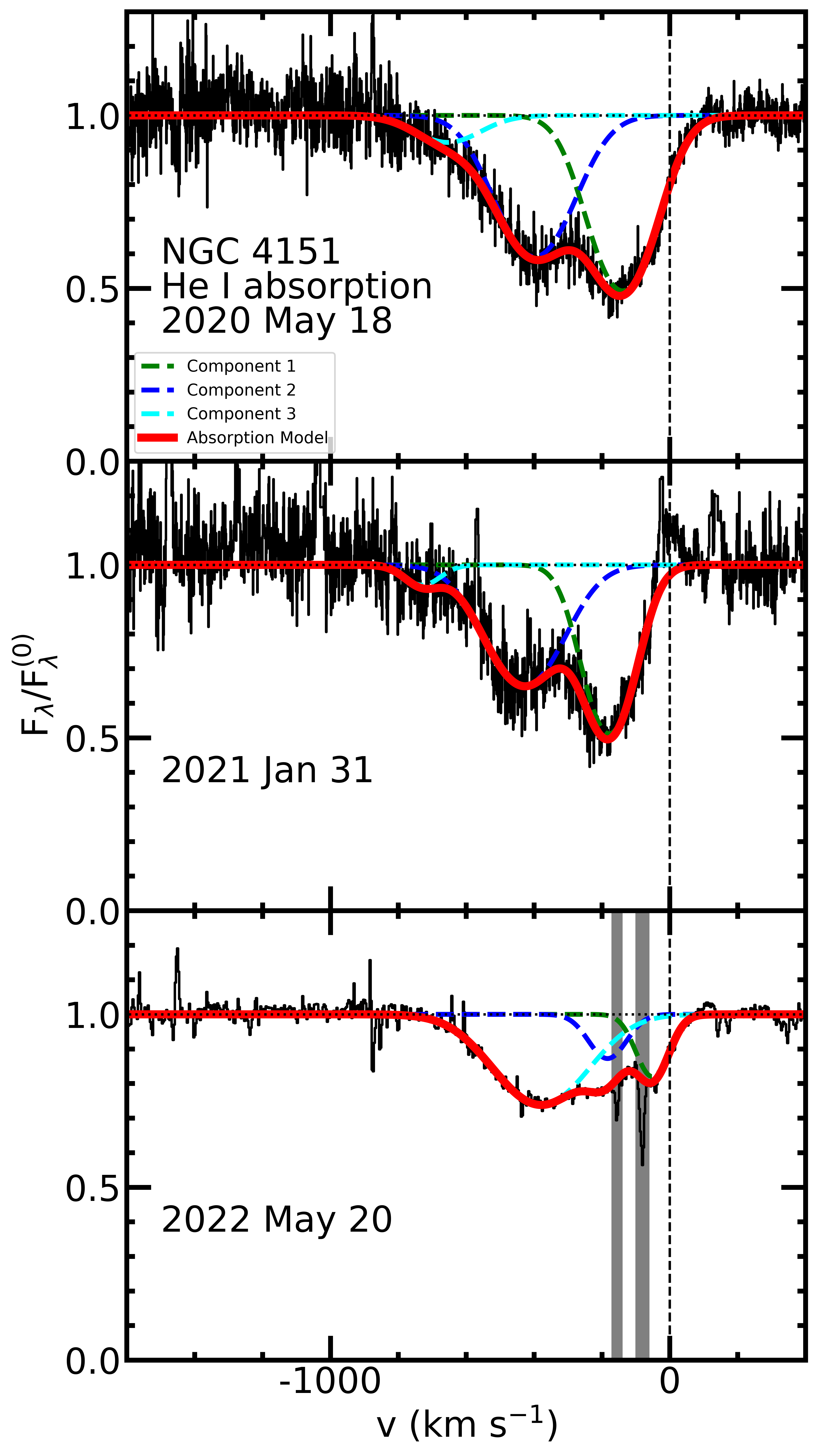}
	\caption{{\itshape Top panel:} spectral decomposition of the He {\scriptsize I} absorption at high resolution in the 2020 epoch. {\itshape Middle panel:} the same for the 2021 epoch. {\itshape Bottom panel:} the same for the 2022 epoch. In the bottom panel, the masks applied to the narrow absorption features mounted on top of the broad He {\scriptsize I} trough ({\it grey bands}) are shown. In all panels, both the emission level normalised to the pseudo-continuum model and the zero-velocity position ({\itshape black dashed lines}) are indicated, and the total absorption model ({\itshape red dashed line}) is plotted along with the single absorption components ({\itshape dot-dashed lines}).}
    \label{fig:ngc4151absfit}
\end{figure}

\begin{table*}[h!t]
    \centering
    \caption{Best-fit parameters of the NGC 4151 NIR spectral features.}
	\label{tab:ngc4151bfp}
    \setlength{\tabcolsep}{10pt}
    \renewcommand{\arraystretch}{1.5}
    \begin{tabular}{lcccr}
	    \hline
		\hline
		Parameter & \multicolumn{3}{c}{Continuum emission} & Units\\
        \cline{2-4}
	       & 2020 & 2021 & 2022 & \\
        \hline
		$\beta_{\rm PL}$ & $-4.9 \pm 1.5$ & $-3.6 \pm 0.1$ & $-3.3 \pm 0.8$ & ---\\
        $T_{\rm BB}^{(1)}$ & $2170 \pm 230$ & $1880 \pm 20$ & $1630 \pm 10$ & K\\
		$T_{\rm BB}^{(2)}$ & $970 \pm 50$ & $870 \pm 10$ & --- & K\\
		\hline
		$\chi^2/\nu_{\rm d.o.f.}$ & 1513.1/1505 & 2073.2/2050 & 390.1/385 & ---\\
		\hline
		\hline
         & \multicolumn{3}{c}{He {\footnotesize I} emission (FWHM$_{\rm NLR} = 220 \pm 20$ km s$^{-1}$)} & \\
        \cline{2-4}
         & 2020 & 2021 & 2022 & \\
        \hline
		FWHM$_{\rm int}$ & $1150 \pm 10$ & $1410 \pm 20$ & $820 \pm 10$ & km s$^{-1}$\\
		FWHM$_{\rm BLR}$ & $4330 \pm 50$ & $4430 \pm 120$ & $3200 \pm 50$ & km s$^{-1}$\\
        $|\Delta v_{\rm int}|$ & $50 \pm 10$ & $\lesssim$30\tablefootmark{*} & $30 \pm 10$ & km s$^{-1}$\\
		$|\Delta v_{\rm BLR}|$ & $840 \pm 40$ & $1030 \pm 100$ & $910 \pm 20$ & km s$^{-1}$\\
		\hline
		$\chi^2/\nu_{\rm d.o.f.}$ & $23569.6/23553$ & $23506.0/23164$ & $5528.5/5508$ & ---\\
		\hline
		\hline
         & \multicolumn{3}{c}{He {\footnotesize I} absorption} & \\
        \cline{2-4}
         & \multicolumn{2}{c}{2020--2021} & 2022 & \\
        \hline
        EW$_{1}$ & \multicolumn{2}{c}{$110 \pm 30$} & $20 \pm 10$ & km s$^{-1}$\\
        $|\Delta v_{1}|$ & \multicolumn{2}{c}{$160 \pm 30$} & $50 \pm 10$ & km s$^{-1}$\\
        FWHM$_{1}$ & \multicolumn{2}{c}{$190 \pm 40$} & $120 \pm 20$ & km s$^{-1}$\\
        EW$_{2}$ & \multicolumn{2}{c}{$110 \pm 20$} & $20 \pm 10$ & km s$^{-1}$\\
        $|\Delta v_{2}|$ & \multicolumn{2}{c}{$420 \pm 30$} & $180 \pm 10$ & km s$^{-1}$\\
        FWHM$_{2}$ & \multicolumn{2}{c}{$260 \pm 20$} & $130 \pm 30$ & km s$^{-1}$\\
		EW$_{3}$ & \multicolumn{2}{c}{$20 \pm 10$} & $90 \pm 10$ & km s$^{-1}$\\
        $|\Delta v_{3}|$ & \multicolumn{2}{c}{$710 \pm 70$} & $380 \pm 10$ & km s$^{-1}$\\
		FWHM$_{3}$ & \multicolumn{2}{c}{$150 \pm 80$} & $310 \pm 20$ & km s$^{-1}$\\
		\hline
        $N^{(1)}_{\rm HeI}(2^3S)$ & \multicolumn{2}{c}{$(0.61 \pm 0.12) \times 10^{13}$} & $(0.12 \pm 0.02) \times 10^{13}$ & cm$^{-2}$\\
		$N^{(2)}_{\rm HeI}(2^3S)$ & \multicolumn{2}{c}{$(0.60 \pm 0.11) \times 10^{13}$} & $(0.10 \pm 0.04) \times 10^{13}$ & cm$^{-2}$\\
		$N^{(3)}_{\rm HeI}(2^3S)$ & \multicolumn{2}{c}{$\lesssim$0.20 $\times 10^{13}$\tablefootmark{*}} & $(0.50 \pm 0.04) \times 10^{13}$ & cm$^{-2}$\\
		$N^{\rm (tot)}_{\rm HeI}(2^3S)$ & \multicolumn{2}{c}{$(1.27 \pm 0.30) \times 10^{13}$} & $(0.72 \pm 0.10) \times 10^{13}$ & cm$^{-2}$\\
		\hline
	\end{tabular}
    \tablefoot{
        The top section reports the best-fit parameters of the NGC 4151 continuum emission components for the analysed epochs of observation, along with the values of the corresponding goodness-of-fit statistical analysis. The central section reports the best-fit velocity parameters of the He {\scriptsize I} $\lambda$10,380 emission features. The bottom section reports the best-fit velocity parameters of the He {\scriptsize I} $\lambda$10,380 absorption features.\\
        \tablefoottext{*}{Upper limit at 95\% confidence level.}
    }
\end{table*}

After this step, we apply the empirical model to the absorption-free intervals of the He {\footnotesize I}$+$Pa$\gamma$ spectral region observed with iSHELL and GIANO-B. In these fits, the 1.08 $\mu$m $\lesssim \lambda \lesssim 1.083$ $\mu$m region is excluded (see Fig. \ref{fig:ngc4151absfit}) due to the presence of the absorption feature, whereas the FWHM of the narrow components is fixed to the value found from the [S {\footnotesize III}] fit performed on the low-resolution spectra\footnote{The correction for the systematic widening due to the instrument resolving power is $\lesssim$6 km s$^{-1}$ in the high-resolution spectra, to be compared with the 150--600 km s$^{-1}$ of the low-resolution ones.} The level of the continuum emission is derived by rescaling the result of the low-resolution spectral modeling by a constant factor, set free to vary in each epoch. In performing the fitting procedure, we note that the full emission profile cannot be precisely recovered unless we include a further intermediate-velocity Gaussian component with FWHM $\sim 1000$ km s$^{-1}$ at all epochs. After having obtained best-fit models for all the three epochs separately, we rescale all spectra and models to the integrated flux of the recovered narrow He {\footnotesize I} emission of the 2020 epoch. The resulting best fits are shown in Fig. \ref{fig:ngc4151he1}; the FWHMs of both broad and intermediate components are reported in Tab. \ref{tab:ngc4151bfp}. The discrepancy at $>$95\% confidence level that insists on the 2022 BLR and intermediate FWHMs with respect to the values obtained for the previous epochs can be ascribed to the custom GIANO-B data reduction performed to recover the broad He {\footnotesize I} emission profile (see Sect. \ref{sec:obsred}).

\subsection{Decomposition of the He {\footnotesize I} absorption}\label{sec:absdec}

\begin{figure*}[htbp]
    \centering
	\begin{minipage}{.33\textwidth}
	\includegraphics[width=\linewidth]{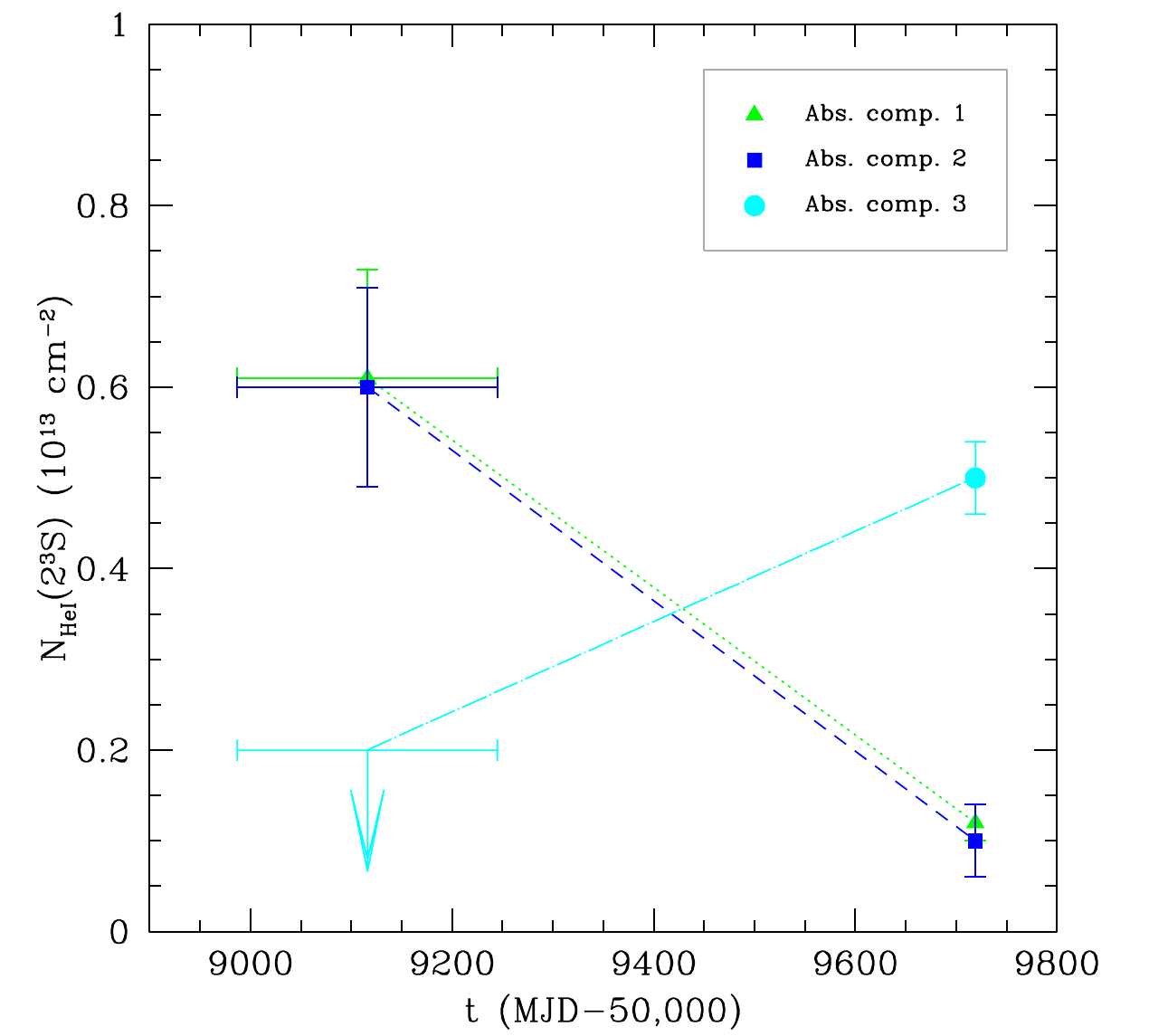}
	\end{minipage}
	\begin{minipage}{.33\textwidth}
	\includegraphics[width=\linewidth]{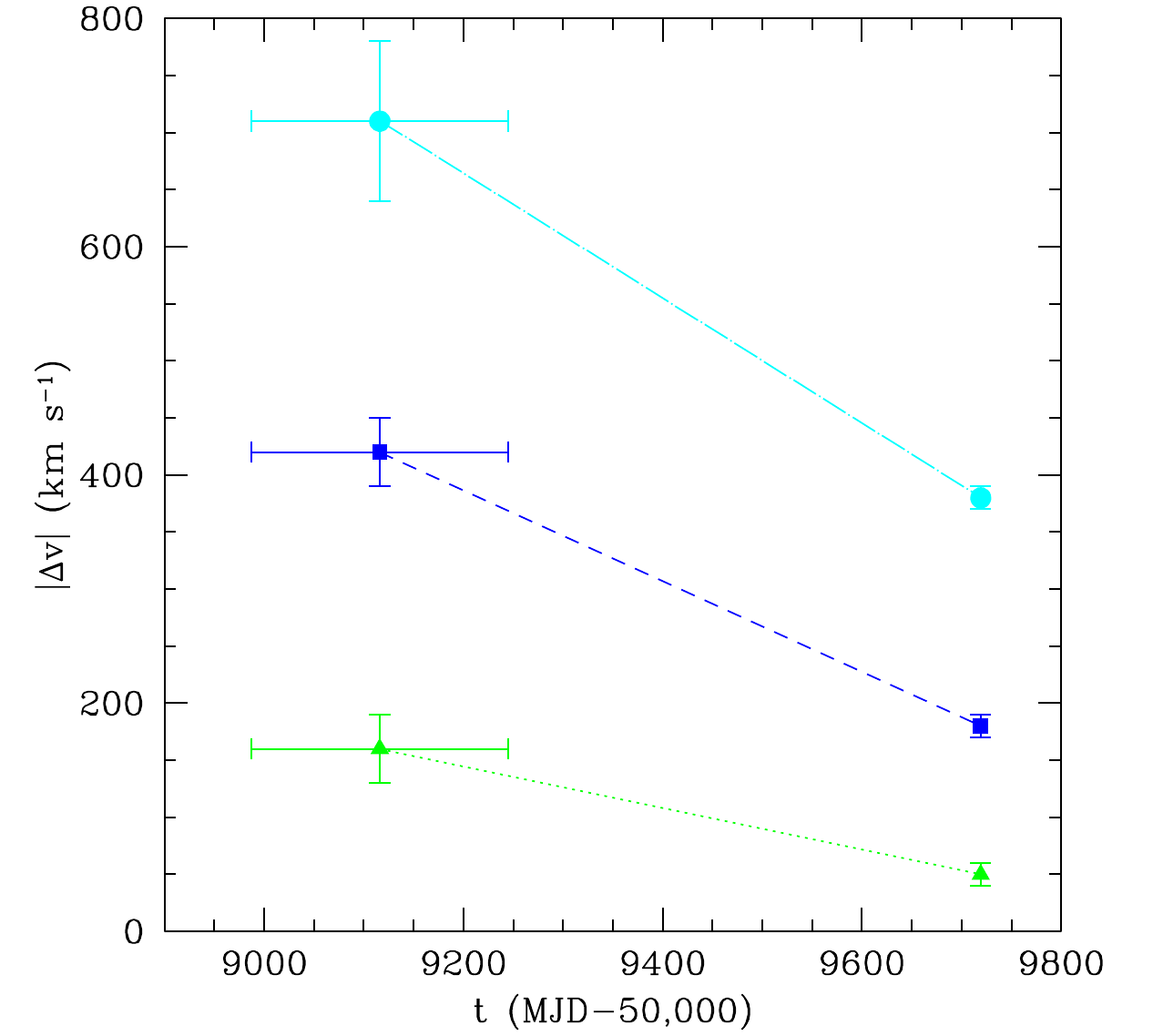}
	\end{minipage}
    \begin{minipage}{.33\textwidth}
	\includegraphics[width=\linewidth]{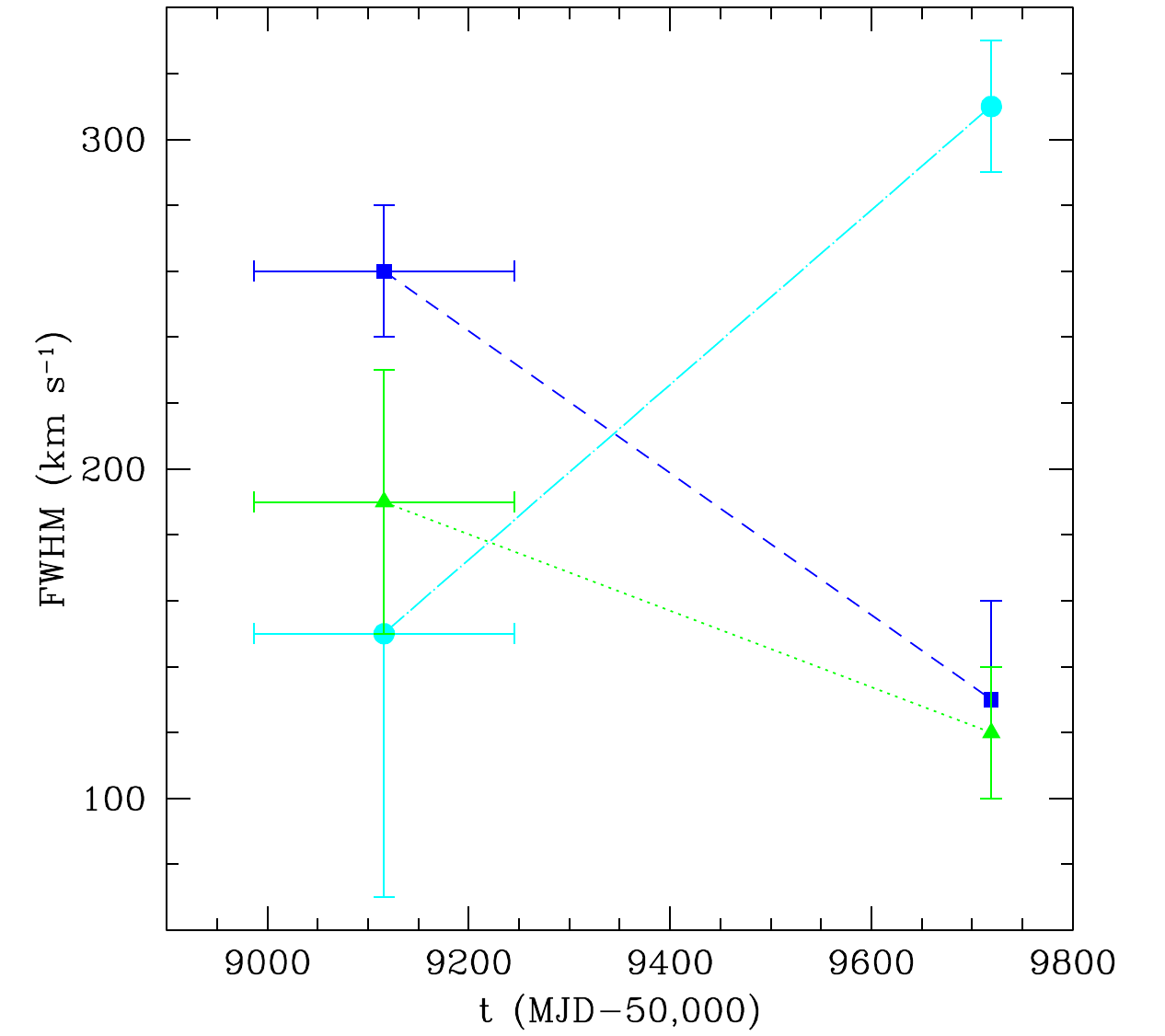}
	\end{minipage}
    \caption{Time variability of the NGC 4151 He {\scriptsize I} $\lambda$10,830 absorption parameters. {\itshape Left panel:} column density of neutral helium atoms in the $2^3${\it S} state. {\itshape Middle panel:} velocity blueshifts. {\itshape Right panel:} FWHMs of each absorption component. In all panels, the three components of the He {\scriptsize I} absorption system are identified by the same colors and symbols ({\itshape see legend}).}
    \label{fig:abspars}
\end{figure*}

After having normalised the observed fluxes to the respective best fits for both epochs, we proceed to model the He {\footnotesize I} absorption feature of NGC 4151. To this end, owing to Eq. \ref{eqn:absfeat}, we assume that the absorbed flux $F(v)$ in the velocity space can be described as \citep[e.g.,][]{ham99}:
\begin{equation}\label{eqn:absflux}
    F(v) = F_0(v) \times e^{-\tau_v}
\end{equation}
where $F_0(v)$ is the unabsorbed flux and $\tau_v$ is the optical depth of the absorption. At a first order, we approximated $\tau_v$ with a sum of Gaussian profiles:
\begin{equation}\label{eqn:optdep}
    \tau_v = \sum_{i=1}^{N} \tau_v^{(i)} = \sum_{i=1}^{N} A_i \times \exp{\left[
    -\frac{\left(
    v - \Delta v_i
    \right)^2}{2 \sigma_i^2}
    \right]}
\end{equation}
with $A_i$, $\Delta v_i$ and $\sigma_i$ the absorption strength, velocity shift and velocity dispersion of each feature, respectively. Such quantities are connected to the corresponding equivalent widths EW$_i$ by:
\begin{equation}\label{eqn:absstr}
    {\rm EW}_i = \int_{-\infty}^{+\infty} \left[
    1 - e^{-\tau_v^{(i)}}
    \right] dv
\end{equation}

We then perform the fit of the He {\footnotesize I} absorbed spectral region by progressively including additional profiles, until a good agreement with the data is reached by the means of both visual inspection and statistical significance. The best fit is obtained for $N = 3$, i.e. decomposing the total absorption into three components at distinct velocities. In doing this, we find that the best-fit parameters -- EWs, velocity shifts, FWHMs -- in the 2020 and 2021 epochs are compatible for each component within the respective 2$\sigma$ errors: this points at a substantial stability of the outflow structure and dynamics over those epochs. Therefore, we describe the He {\footnotesize I} absorption in these two epochs by computing the uncertainty-weighted averages of such parameters. In contrast, a visual inspection of the He {\footnotesize I} absorption in the 2022 epoch reveals a radically different trough shape with respect to the previous epochs: for this reason, we study this epoch separately from the other two. In addition, the fit to the 2022 He {\footnotesize I} broad absorption requires to mask two further narrow components that appear at $|\Delta v| \lesssim 200$ km s$^{-1}$, whose study is postponed to future work. We report the values of EW, blueshift and FWHM obtained in this way in Tab. \ref{tab:ngc4151bfp}; in Fig. \ref{fig:ngc4151absfit}, we show the resulting modeling of the He {\footnotesize I} absorption system.

Finally, we compute the outflow column density $N_{\rm HeI}^{(i)}$ associated with each absorption component by integrating the corresponding optical depth in the velocity space \citep{sav91,wil16}:
\begin{equation}\label{eqn:nion}
    N_{\rm HeI}^{(i)} = \frac{m_e c}{\pi f \lambda_0 q^2} \int_{-\infty}^{+\infty} \tau(v) dv
\end{equation}
with $m_e$ the electron mass, $c$ the speed of light, $f$ the He {\footnotesize I} oscillator strength \citep{wei67}, $\lambda_0$ the laboratory rest-frame wavelength and $q$ the elementary charge. Taking $f = 0.6257$ for the total $2^3S$ transitions of the He {\footnotesize I} \citep{spi98}, we calculate $N_{\rm HeI}^{(i)}$ for each absorber at all epochs. For the highest-velocity component, the associated column density is compatible with zero in both epochs after having propagated the best-fit parameter uncertainties in Eq. \ref{eqn:nion}. Since also all of the $N_{\rm HeI}^{(i)}$ are compatible within the respective 1$\sigma$ uncertainties between the 2020 and 2021 epochs, we take the weighted averages over these two epochs for them too (see Tab. \ref{tab:ngc4151bfp}). Overall, a decreasing trend in the total absorber density is present between 2020--2021 and 2022, where both the shape and the physical parameters of the He {\footnotesize I} absorption undergo relevant changes (see Fig. \ref{fig:ngc4151absfit}). The most striking feature among these variations is the increase in absorption strength and velocity of the fastest component, which points at changes in the outflow structure.

To quantify these effects, we study for each component the time evolution of the absorption parameters -- $2^3S$ He {\footnotesize I} column densities, absolute values of the velocity blueshifts $|\Delta v|$, and FWHMs. For each parameter, we construct the time variability curve between 2020 and 2022; the results are reported in Fig. \ref{fig:abspars}. A general slowdown of all the absorption components by a factor of $\sim$2 at $>$99\% confidence level accompanies the reversal in both $N_{\rm HeI}$ and FWHM of the high-velocity component with respect to the lower-velocity ones. Changes in the He {\footnotesize I} absorption structure were already detected by \citet[see their figure 8]{wil16}, with blue-shifts spanning from $\sim$100 km s$^{-1}$ in their Epoch 2 to $\sim$800 km s$^{-1}$ in their Epoch 3, i.e. over a time interval of $\sim$2 yr; furthermore, the bulk outflow velocity monotonically decreases down to $\sim$400 km s$^{-1}$ over their following two epochs. We highlight that such a decomposition of the He {\footnotesize I} absorption profile is purely empirical, and is adopted with the main purpose of studying the trough properties and variability through a smooth line profile that is free of noise or intervening spurious components (see Fig. \ref{fig:abspars}), and allows for an analytic calculation of the associated physical quantities. In the following, we thus only discuss the global absorption properties rather than further discriminating among individual components of the He {\footnotesize I} trough.

\section{{\it Swift} X-ray data}\label{sec:xraydat}

\begin{figure*}
    \centering
    \sidecaption
    \includegraphics[width=12cm]{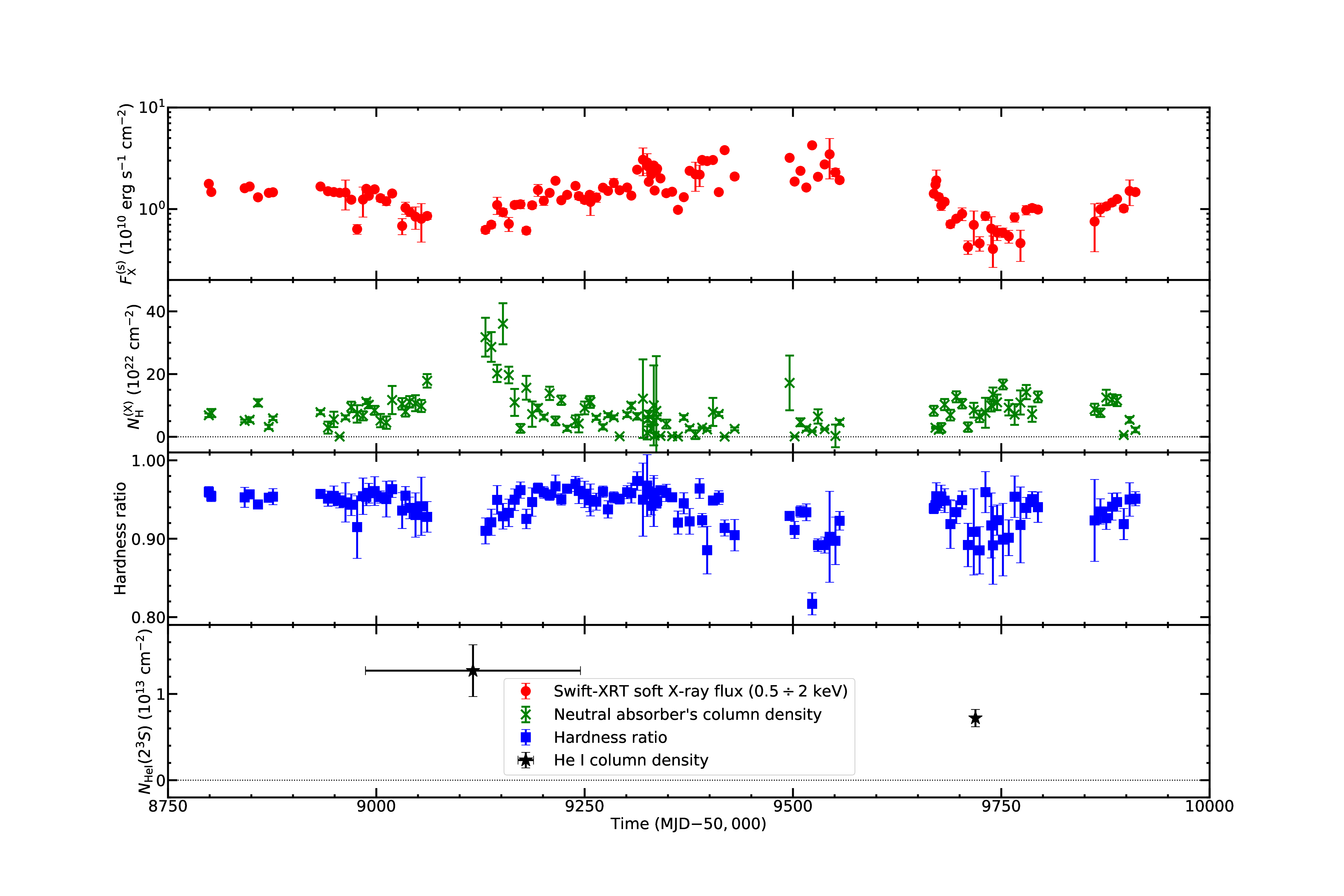}
    \caption{Multi-epoch evolution of the NGC 4151 X-ray parameters compared to the He {\scriptsize I} column density, along with the corresponding 1$\sigma$ uncertainties. {\it Top panel:} soft X-ray flux $F_{\rm X}^{\rm (s)}$ (0.5--2 keV, {\it red points}). {\it Upper mid panel:} column density $N_{\rm H}^{\rm (X)}$ of the X-ray neutral absorber ({\it green crosses}). {\it Lower mid panel:} hardness ratio of the X-ray spectrum ({\it blue squares}. {\it Bottom panel:} column density $N_{\rm HeI}^{\rm (tot)}(2^3{\rm S})$ of the He {\scriptsize I} absorber ({\it black stars}). In the column density panels, the zero level {\it dotted line} is indicated.}
    \label{fig:mwl}
\end{figure*}

It is presently known that quasars with gas outflows often exhibit multiple absorption troughs in several bands \citep[e.g.,][]{cic18}; such multi-phase outflows are ubiquitous across the cosmic time, being found from the local Universe \citep[e.g., Mrk 231 at $z \sim 0.04$;][]{fer10} to epochs much earlier than the quasar activity peak at $z \sim 2$ \citep[e.g., SDSS J153830.55$+$085517.0 at $z \sim 3.6$;][]{vie22}, and can consistently vary together in response to changes in the structure and/or physical properties of the absorbing gas \citep[e.g.][]{}. NGC 4151 is known to show absorption in other bands, such as in the X-rays \citep[][]{Beuchert2017,Gianolli2023}, in which it has been subject to extensive observing campaigns with e.g. the X-Ray Telescope (XRT) on board of the {\it Swift} satellite \citep[{\it Swift}-XRT;][]{bur00}. We therefore performed the analysis of the complete public {\it Swift}-XRT dataset of NGC 4151 in search of indications of common variability properties between the He {\footnotesize I} absorption, the X-ray absorber and the X-ray ionizing flux.

To seek for a connection between the X-ray and IR data of NGC 4151, we collect all the available {\it Swift}-XRT observation available between April 2019 and December 2023, for a total of 165 epochs. The high-level science products have been obtained using the standard pipelines \texttt{xrtpipeline} and \texttt{xrtproducts}\footnote{See \url{https://swift.gsfc.nasa.gov/analysis/xrt_swguide_v1_2.pdf}.}. To extract the source spectrum and take into account any pile-up issues, we use an extracting region with a variable shape. A circle or an annulus, respectively, are adopted in the cases of a source rate $<$0.6 s$^{-1}$ or $>$0.6 s$^{-1}$. The inner radius of the region is determined on the basis of the observed count rate as per table 2 of \citet{Middei2022}. Regardless of its circular or annular shape, the source extracting region always has a (outer) radius of 50$''$. For the background, we use an annulus centered on the source with a fixed gap of 25 pixels ($\sim$60$''$) between the inner and outer radii. The obtained spectra have been subsequently binned, requiring at least 5 counts per bin.

For each of the 165 observations, we fit the corresponding spectrum with {\footnotesize XSPEC} \citep{Arnaud1996}, adopting the following model:
\begin{equation}
    \mbox{\texttt{tbabs}}_{\rm MW} \times \mbox{\texttt{ztbabs}} \times \mbox{\texttt{zxipcf}} \times \mbox{\texttt{po}}\nonumber
\end{equation}
The first component accounts for the Milky Way hydrogen column $N_{\rm H}$, whereas the second one accounts for the local neutral absorption $N_{\rm H}^{\rm (X)}$ commonly observed in NGC 4151. Then, \texttt{zxipcf} is used to reproduce the WAs while a PL emission is assumed for the primary continuum emission. In the fits, we keep fixed the Milky Way $\rm N_{\rm H} = 2.1\times10^{20}~cm^{-2}$ \citep[][]{HI4PI2016}, while we leave $N_{\rm H}^{\rm (X)}$ free to vary. For the warm absorbing matter we compute the column density, the ionisation level and the covering fraction in each observation. Finally, only the normalization of the the primary continuum is fitted, fixing the PL photon index $\Gamma$ to a standard value of 1.8 \citep[see e.g.][]{Piconcelli2005,Bianchi2009,Matzeu2023} to mitigate the known degeneracy between the source spectral shape and the column density of the absorbers.

We present the main results of the X-ray data analysis -- namely the soft photon flux $F_{\rm X}^{\rm (s)}$, the hardness ratio and the X-ray neutral absorber's column density $N_{\rm H}^{\rm (X)}$ -- in Fig. \ref{fig:mwl} as a function of the observing epoch. Since until MJD $\sim 58,200$ the {\it Swift}-XRT coverage of NGC 4151 is sparse, we restrict our inspection to the most populated time range MJD = 58,200--60,200; this allows us to directly compare the behaviour of the NGC 4151 X-ray properties with the He {\footnotesize I} absorbed epochs studied in this paper. A visual inspection of Fig. \ref{fig:mwl} reveals that, around MJD $\sim 59,200$, $N_{\rm H}^{\rm (X)}$ exhibits a peak -- with an increase in intensity by a factor of $\sim$7 -- simultaneous to the two epochs with the most intense He {\footnotesize I} absorption; subsequently, when the $N_{\rm H}^{\rm (X)}$ strength has already decreased, a peak in $F_{\rm X}^{\rm (s)}$ (MJD $\sim 59,500$) -- with an increase by a factor of $\sim$40 -- happens just before the observed decrease of the He {\footnotesize I} column density. These changes in the X-ray flux are consistent with the variations found in the NGC 4151 hardness ratio over the same time period. Though interesting, however, such temporal coincidences hint at a potential correlation between the He {\footnotesize I} absorber status and the X-ray activity of NGC 4151 with a too marginal confidence to allow reaching firm conclusions.

\begin{figure}
    \centering
    \resizebox{\hsize}{!}{
    \includegraphics[angle=-90]{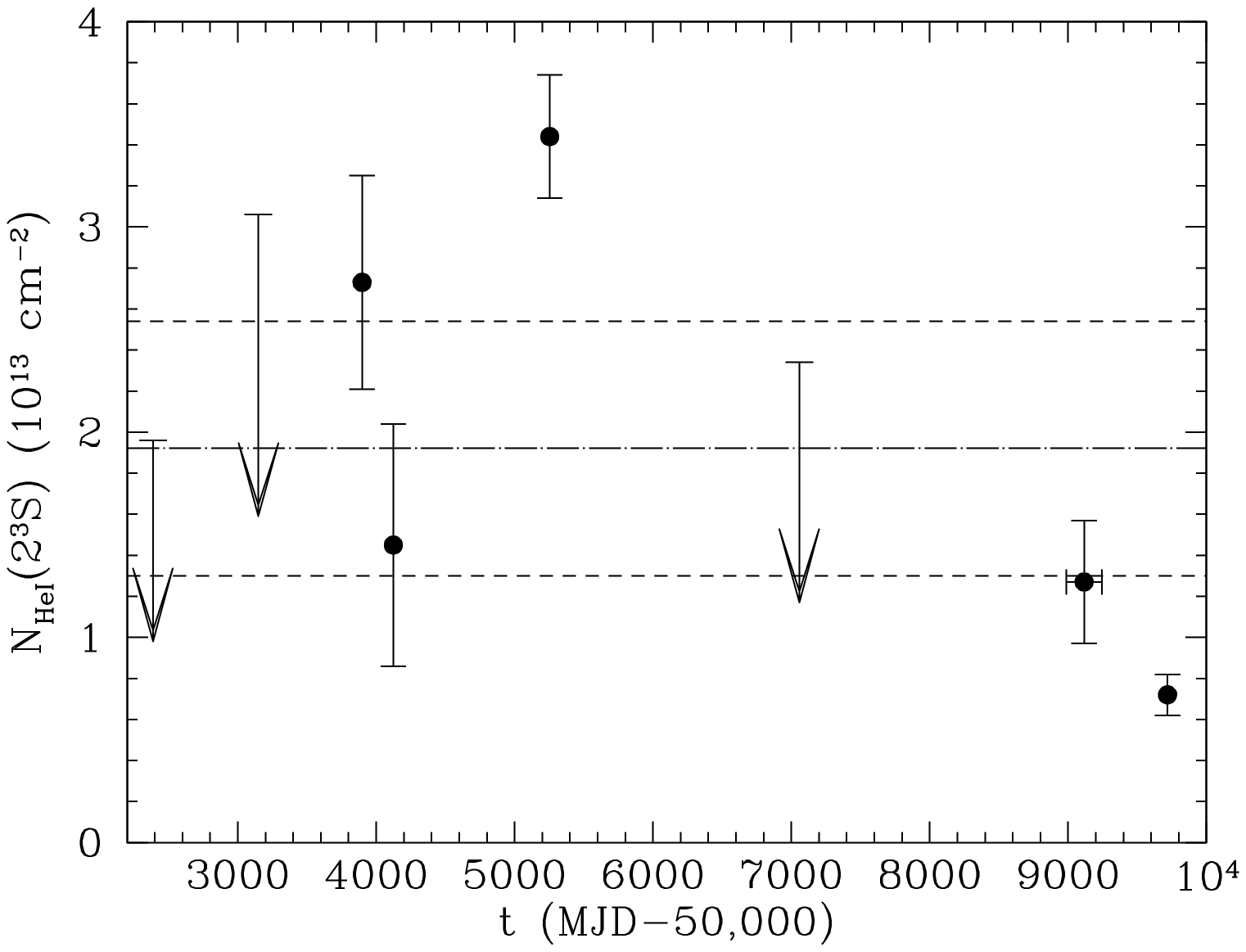}
    }
    \caption{Evolution of the He {\scriptsize I} column density over 20 years of IR spectroscopic observations of NGC 4151. Epochs before MJD $= 58,000$ are taken from \citet{wil16}. Upper limits ({\it black arrows}) are computed at 2$\sigma$ confidence level. The average of the significant He {\scriptsize I} measurements ({\it dot-dashed line}), along with its 1$\sigma$ uncertainty ({\it dashed line}), is indicated.}
    \label{fig:trend}
\end{figure}

\section{Discussion}\label{sec:disc}

The He {\footnotesize I} absorption trough of NGC 4151 has clearly undergone a change in shape and intensity over a 1-year timescale, with a general decreasing trend in both strength and velocity spread. This effect is well documented in the literature, especially in the case of AGN with BALs \citep[BAL QSOs;][]{bar93} for which both single-objects \citep{bar92,tre13} and ensemble studies \citep{lun07,gib08,gib10,cap11,cap12,cap13,fil13} have provided information on the time variability of such features on several scales (from $\sim$10$^{-2}$ to $\sim$10 years). In some extreme cases, the appearance \citep[e.g.,][]{kro10,vie22} or disappearance \citep[e.g.,][]{hal11} of BALs in quasar spectra has been documented. In the following, we discuss the He {\footnotesize I} absorption variability in connection with the activity of the central engine and derive estimates of the physical properties of the parent outflow.

\subsection{He {\footnotesize I} absorption variability}\label{sec:absvar}

To investigate the driving mechanism of the He {\footnotesize I} absorption variability in NGC 4151, we construct the temporal evolution of the $N_{\rm HeI}(2^3S)$ parameter by joining the measurements performed by \citet{wil16} with those presented in Tab. \ref{tab:ngc4151bfp}; in this way, we obtain a ``dark curve'' of the NGC 4151 He {\footnotesize I} absorption extended over $\sim$20 years\footnote{The same duration of the time interval holds for both the observer's and the rest frame, given the NGC 4151 low redshift $z = 0.0033$.}. We show the absorption trend in Fig. \ref{fig:trend}; a visual inspection reveals He {\footnotesize I} obscured epochs alternated to unabsorbed periods. To quantify the significance of these variations in the He {\footnotesize I} column density over time, we perform a Cox regression test \citep{cox72} on the time series shown in Fig. \ref{fig:trend}. This method allows to take into account both data uncertainties and upper limits, similarly to other common-use statistical methods for analysing left-censored astronomical data \citep[e.g.,][]{fei85,iso86}. Assuming a constant $N_{\rm HeI}(2^3S) = (1.92 \pm 0.62) \times 10^{-13}$ cm$^{-2}$ -- corresponding to the average of the significant He {\footnotesize I} column density values -- over the MJD 52,200--60,000 time period as the null hypothesis, the test yields an associated probability of 0.04, that implies a 96\% ($>$2$\sigma$) significance of the $N_{\rm HeI}(2^3S)$ variability.

The main driving mechanism of BAL variability is still a subject of debate: to date, there are in fact no decisive observations that may allow to discriminate among the proposed scenarios to explain this phenomenon, that is commonly attributed to a combination of variable ionizing flux and motion of gas into and out of the line of sight \citep[e.g.,][]{cre03}. The only other object that is known to have a multi-phase outflow visible from the X-rays to the IR wavelengths is the nearby AGN NGC 5548 \citep[e.g.,][]{ede99,kaa00}. \citet{wil21} discovered in fact that this active galaxy is also hosting a He {\footnotesize I} absorption trough whose variations are directly influenced by changes in the soft X-ray flux that also triggers the variability of the WA component. In addition, they found that some of the NGC 5548 He {\footnotesize I} absorption components were connected to an obscuration event happened in 2013. In this way, they were able to describe the inner structure of NGC 5548 with the accretion disk wind model by \citet{deh19a,deh19b}, in which the absorption variability is partly due to changes in the density of the obscurer and partly responding to X-ray flux variations.

Based on the hints provided by the analysis of the NGC 4151 X-ray data, it is plausible that the NGC 4151 He {\footnotesize I} absorption variability is partially connected with changes in the X-ray properties (see Sect. \ref{sec:xraydat}). In this scenario, periods of increasing X-ray flux result in decreasing absorption strengths due to the higher ionisation of atoms; conversely, low-emission states allow the outflowing gas to recombine, giving rise to stronger absorption features. Clearly, the evidence of the association between the NGC 4151 X-ray and absorption variability presented here is weak, and only tentatively based on visual inspections of time series coming from data analyses of IR and X-ray spectra. The application of quantitative methods to estimate the likelihood of such an association, such as the calculation of cross-correlation functions \citep[e.g.,][]{gas87,ede88,ale97} between the studied time series cannot be considered at the moment, given the paucity and temporal scatter of He {\footnotesize I} absorption measurements.

\subsection{Physical properties of the parent outflow}\label{sec:outphys}

Having derived in Sect. \ref{sec:specdec} the velocity and density properties of each He {\footnotesize I} absorption component (see Tab. \ref{tab:ngc4151bfp}), we proceed to use them to estimate some physical properties of the parent outflow. Measurements of quantities such as the outflowing mass rate $\dot{M}_{\rm out}$, kinetic power $\dot{E}_{\rm kin}$ and momentum rate $\dot{P}_{\rm out}$ are traditionally performed on AGN exhibiting the presence of ionised winds detected in emission in their spectra \citep[see e.g.][and refs. therein]{sat21}. However, in recent years similar calculations have started also for outflows seen as absorption troughs in AGN and starburst galaxies \citep[e.g.,][]{xu23,deh25}. To get a rough idea of the energy budget involved in the He {\footnotesize I} outflow of NGC 4151, we follow the approach of \citet{deh25} by computing $\dot{M}_{\rm out}$ as:
\begin{equation}\label{eqn:outmrate}
    \dot{M}_{\rm out} \simeq 4 \pi f_{\rm int} r_{\rm out} N_{\rm H} \mu m_{\rm p} v_{\rm max}
\end{equation}
where $v_{\rm max}$ is the maximum outflow velocity, $m_{\rm p}$ the proton mass, $\mu = 1.4$ the average atomic mass per proton, $N_{\rm H}$ the hydrogen column density, $r_{\rm out}$ the outflow spatial extension and $f_{\rm int}$ the intrinsic covering fraction of the outflow. This quantity allows in turn to determine the outflow kinetic power $\dot{E}_{\rm kin}$ and momentum rate $\dot{p}_{\rm out}$ as \citep{sat21,deh25}:
\begin{eqnarray}\label{eqn:outepr}
    \dot{E}_{\rm kin} \simeq \frac{1}{2}\dot{M}_{\rm out} v_{\rm max}^2\\
    \dot{p}_{\rm out} \simeq \dot{M}_{\rm out} v_{\rm max}
\end{eqnarray}

\begin{table}[h!t]
    \centering
    \caption{Summary of the relevant physical properties of the He {\scriptsize I} outflow detected in the NGC 4151 IR spectrum.}
    \label{tab:outpars}
    \resizebox{\columnwidth}{!}{
    {\setstretch{1.5}
    \begin{tabular}{lcr}
    \hline
    \hline
        Quantity & Value & Units \\
    \hline
        $v_{\rm max}$ & $840 \pm 140$ & km s$^{-1}$\\
        $r_{\rm out}$ & $6.7^{+2.9}_{-0.7}$ & pc\\
        $\dot{M}_{\rm out}$ & $\left( 1.8^{+0.9}_{-1.1} \right) \times 10^{-3}$ & M$_\odot$ yr$^{-1}$\\
        $\dot{E}_{\rm kin}$ & $\left( 4.0^{+4.2}_{-2.9} \right) \times 10^{38}$ & erg s$^{-1}$\\
        $\dot{p}_{\rm out}$ & $\left( 9.6^{+7.2}_{-6.5} \right) \times 10^{30}$ & erg cm$^{-1}$\\
    \hline
    \end{tabular}
    }
    }
\end{table}

Taking $v_{\rm max} = |\Delta v_3| + 2 \sigma_3$ \citep{bis17} equal to $840 \pm 140$ km s$^{-1}$ from our spectral analysis (see Tab. \ref{tab:ngc4151bfp}), with $\sigma_3 = {\rm FWHM}_3 / 2 \sqrt{2 \ln{2}}$, the only unknown quantities in Eq. \ref{eqn:outmrate} are $N_{\rm H}$, $f_{\rm int}$ and $r_{\rm out}$. The second is usually assumed to be equal to the overall detected fraction of BAL QSOs, which is found to be $\approx$0.2 at all cosmic times \citep[e.g.,][]{hew03,gib09,all11,dai12}; for the latter, we estimate it as done by \citet[see their eq. 3]{lam03}:
\begin{equation}\label{eqn:outrad}
    \frac{r_{\rm out}}{\rm cm} \simeq 4 \times 10^{16} M_7^{1/5} \left(
        \frac{L_{42} t_{\rm days}}{N_{22} U}
    \right)^{2/5}
\end{equation}
Here, $M_7$ is the SMBH mass in units of $10^7$ M$_\odot$, $L_{42}$ the ionising luminosity in units of $10^{42}$ erg s$^{-1}$, $t_{\rm days}$ the variability timescale of the absorption expressed in days, $N_{22}$ the hydrogen column density in units of $10^{22}$ cm$^{-2}$, and $U$ the ionisation parameter.

We estimate $L_{42}$ over the range of unobserved ionising energies following the procedure detailed in \citet{gri15}, calibrating the synthetic quasar SED of \citet{dun10} to the NGC 4151 bolometric luminosity $L_{\rm bol}$ from its AGN activity. From the {\it Swift}-XRT X-ray data, we derive $F_{\rm X}^{\rm (s)} \sim 3 \times 10^{-12}$ erg s$^{-1}$ cm$^{-2}$ for the epochs free of major flares (see Fig. \ref{fig:mwl}), corresponding to a soft X-ray luminosity $L_{\rm X}^{\rm (s)} \sim 7.2 \times 10^{40}$ erg s$^{-1}$. We use this value to solve equation 21 by \citet{mar04} for $L_{\rm bol}$:
\begin{equation}\label{eqn:lbol}
    \log{\left[
    \frac{L_{\rm bol}}{L_{\rm X}^{\rm (s)}}
    \right]} = 1.65 + 0.22 \mathcal{L} + 0.012 \mathcal{L}^2 - 0.0015 \mathcal{L}^3
\end{equation}
where $\mathcal{L} = \log{(L_{\rm bol}/{\rm L}_\odot)} - 12$. In turn, we get $L_{\rm bol} \sim 8.9 \times 10^{41}$ erg s$^{-1}$ and an ionising luminosity normalisation $L_{911} \sim 5.5 \times 10^{25}$ erg s$^{-1}$ Hz$^{-1}$, adopting the bolometric correction at 911 \AA\ by \citet[see their figure 12]{ric06}. Integrating the synthetic SED over the range of ionising frequencies, we are thus able to estimate an NGC 4151 ionising luminosity $L_{\rm ion} \sim 3.2 \times 10^{41}$ erg s$^{-1}$, corresponding to $L_{42} \sim 0.32$.

The outflow helium, hydrogen and electron densities $n_{{\rm He}^+}$, $n_{\rm H}$ and $n_{\rm e} \sim 1.2 n_{\rm H} \sim 13.3 n_{{\rm He}^+}$ \citep{ost06,deh25} for a typical plasma composed by $\sim$90\% of hydrogen and $\sim$9\% of helium and metals \citep{xu23}, can be related to the number density of helium atoms in the $2^3S$ state via equation 4 by \citet{fer86}:
\begin{equation}\label{eqn:23she}
    \frac{n_{2^3S}}{n_{{\rm He}^+}} = \frac{6.13 \times 10^{-6} T_4^{-0.94}}{1 + 3390 T_4^{-0.21}/n_{\rm e}}
\end{equation}
with $T_4$ the plasma temperature normalised to $10^4$ K. Replacing $n_{{\rm He}^+}$ with its relation to $n_{\rm H}$ and inverting this equation for the latter, we find:
\begin{equation}\label{eqn:edens}
    N_{\rm H} \approx 1.8 \times 10^6 T_4^{0.94} \frac{N_{2^3S}}{r_{\rm out}}
\end{equation}
once having removed the negligible terms and having further assumed $n \sim N/r_{\rm out}$ for all densities. Adopting a typical Seyfert outflow temperature $T \sim 14,000$ K \citep[e.g.,][]{deh24} -- i.e. $T_4 \sim 1.4$ -- and $N_{2^3S} = (0.9 \pm 0.5) \times 10^{13}$ cm$^{-2}$ -- computed by averaging the values reported in Tab. \ref{tab:ngc4151bfp} -- we solve Eq. \ref{eqn:outrad} with the adoption of $t_{\rm days} = 601 \pm 129$ rest-frame days from the time difference between the GIANO-B and the IRTF observations (see Tab. \ref{tab:jobs}), a mean $U \sim 0.04$ \citep{ale99} and a SMBH mass $M_{\rm BH} = (1.7 \pm 0.4) \times 10^7$ M$_\odot$ \citep{ben22} for NGC 4151. In this way, we finally obtain $r_{\rm out} \simeq 6.8^{+2.9}_{-0.7}$ pc, corresponding to $N_H \approx (1.1 \pm 0.8) \times 10^{19}$ cm$^{-2}$ -- i.e. $N_{22} \approx (1.1 \pm 0.8) \times 10^{-3}$. Such a value is in agreement with the typical location at $\sim$pc scales of AGN absorbers found in the literature for similar targets \citep[see e.g.][and refs. therein]{tom12}.

Through Eq. \ref{eqn:outmrate}, we can now compute a value of $\dot{M}_{\rm out} \simeq \left( 1.8^{+0.9}_{-1.1} \right) \times 10^{-3}$ M$_\odot$ yr$^{-1}$. This yields in turn $\dot{E}_{\rm kin} \simeq \left( 4.0^{+4.2}_{-2.9} \right) \times 10^{38}$ erg s$^{-1}$ and $\dot{p}_{\rm out} \simeq \left( 9.6^{+7.2}_{-6.5} \right) \times 10^{30}$ erg cm$^{-1}$. We summarise our findings on the NGC 4151 He {\footnotesize I} outflow energetics in Tab. \ref{tab:outpars}. Overall, the estimated quantities are in line with similar values found in Seyfert galaxies and quasars of comparable energetics exhibiting spectral outflows seen in both absorption \citep[see][and refs. therein]{deh25} and emission \citep[see][and refs. therein]{sat21}. It is to be noted that the $\dot{E}_{\rm kin}$ reported in Tab. \ref{tab:outpars} is negligible ($\ll$0.1\%) if compared to the NGC 4151 Eddington luminosity $L_{\rm Edd} \sim 1.9 \times 10^{45}$ erg s$^{-1}$, as also found in similar sources \citep[e.g.,][]{blu05}; given that a minimal ratio $\dot{E}_{\rm kin}/L_{\rm Edd} \approx 0.5$\% is required for an outflow to significantly contribute to the AGN feedback \citep{hop10}, the ionised wind in NGC 4151 is too weak to have a considerable impact on the gas reservoir of the host galaxy, even in the case of taking into account the systematic uncertainties insisting on the adopted equations that exert biases at the level of 1--2 dex \citep{bis17,sat21}.

\section{Summary and conclusions}\label{sec:conc}

In this paper, we have presented a multi-wavelength study of the inner engine of the Seyfert galaxy NGC 4151, taking advantage of high-resolution NIR {\'e}chelle spectroscopy carried out with the iSHELL and GIANO-B instruments mounted at the {\it IRTF} and {\it TNG} telescopes, respectively. The $>$50,000 resolving power of these spectrographs has allowed a detailed decomposition over a 3-year period of the absorbing trough associated with the He {\footnotesize I} $\lambda$10,830 emission, that was already the subject of one of the few existing works on this topic \citep{wil16}. To infer some connection of this absorption feature with the AGN ionising flux, we also performed the analysis of the NGC 4151 X-ray emission over the same time span, taking advantage of the public {\it Swift}-XRT data. We summarise our main findings as follows:

\begin{enumerate}    
    \item the He {\footnotesize I} absorption system can be empirically decomposed into three components, that highlight the variation of the trough structure over time -- velocity shifts decreasing by a factor of $\sim$2 in $\sim$450 days from MJD $\sim 59,250$ to MJD $\sim 59,700$ (see Fig. \ref{fig:abspars}), accompanied by changes in the overall absorption profile -- and allow the measurement of outflow velocities up to $\sim$800 km s$^{-1}$ and total column densities of $\sim$10$^{13}$ cm$^{-2}$; in all epochs, such values are lower than the column densities found by \citet{wil16} for the total absorption ($\lesssim$3 $\times 10^{13}$ cm$^{-2}$) during the NGC 4151 obscured epochs around MJD $\sim$ 55,000;
    
    \item the investigation of the NGC 4151 X-ray emission revealed that the highest intensity of the He {\footnotesize I} absorption coincides with an increase by a factor of $\sim$4 in the column density of the X-ray neutral absorber (see Fig. \ref{fig:mwl}); similarly, the decrease in $N_{\rm HeI}$ happens after an increase by a factor of $\sim$20 in the X-ray flux, lasting approximately for the duration of the time gap between the second {\it IRTF}/iSHELL observation and the {\it TNG}/GIANO-B epoch;
    
    \item the parent outflow that originates the He {\footnotesize I} absorption trough exhibits properties typical of similar processes at work in AGN \citep[see Tab. \ref{tab:outpars}; e.g.,][]{blu05,tom12}, although it is not powerful enough to efficiently trigger a global AGN feedback \citep[$\dot{E}_{\rm kin}/L_{\rm Edd} \ll 0.1$\%;][]{hop10}.
\end{enumerate}

These findings suggest a scenario in which -- similarly to what happens in BAL QSOs \citep[e.g.,][]{tre13,sat16} and in NGC 5548 \citep{wil21} -- both (part of) the X-ray absorption and the He {\footnotesize I} feature are produced by the same clumpy outflow, whose physical properties change by either through modifications in the outflow structure or by responding to variations of the X-ray flux. The future investigation of the rare He {\footnotesize I} absorption with high-resolution near-IR instruments, also in connection with multi-wavelength observations in other energy bands (e.g., UV/optical, X-rays), will be of extreme interest to infer the properties of the less common cold obscurers with respect to the more ubiquitous ionised outflows; in this framework, more detailed studies of the mechanisms acting behind the NGC 4151 outflows in connection with the AGN activity -- e.g., by modelling the outflow physics with the {\footnotesize CLOUDY} photo-ionisation code \citep{fer98} -- are also needed, as well as a denser monitoring of the NGC 4151 central engine in a multi-wavelength framework to allow for establishing more quantitative correlations among the several observable features -- AGN ionising flux, absorption strength, outflow velocity -- with rigorous techniques for time-series analysis \citep[e.g.,][]{gas87,ede88,ale97}.

\begin{acknowledgements}
    We acknowledge Ennio Poretti (INAF-OAB) for the award of INAF-TNG DDT observing time for NGC 4151 with GIANO-B. We thank prof. Martin J. Ward (University of Durham) for his help with the scientific interpretation of the analysed data, Avet Harutyunyan (INAF-TNG) and Simone Antoniucci (INAF-OAR) for their assistance with the observation preparation and data reduction. We also thank the anonymous referee for their helpful comments. RM acknowledges financial support from the INAF Scientific Directorate. HL acknowledges a Daphne Jackson Fellowship sponsored by the Science and Technology Facilities Council (STFC), UK, and support from STFC grants ST/P000541/1, ST/T000244/1 and ST/X001075/1. HL was the astronomer observing with the Infrared Telescope Facility (IRTF), which is operated by the University of Hawaii under contract 80HQTR24DA010 with the National Aeronautics and Space Administration (NASA). Based on observations made with the Italian {\it Telescopio Nazionale Galileo} (TNG), operated on the island of La Palma by the Fundaci{\'o}n Galileo Galilei of the INAF (Istituto Nazionale di Astrofisica) at the Spanish {\it Observatorio del Roque de los Muchachos} of the Instituto de Astrof{\'i}sica de Canarias. The processed data underlying this work are available on request from the authors. The IRTF raw data are publicly available at the NASA IRTF Archive hosted by the NASA/IPAC Infrared Science Archive (\url{https://irsa.ipac.caltech.edu}). The {\it Swift} data are available to the scientific community through data centres in the USA, Italy, and the UK. Reproduced with permission from Astronomy \& Astrophysics, \textcopyright\ ESO
\end{acknowledgements}

\bibliographystyle{aa}
\bibliography{ngc4151biblio}

\end{document}